\documentclass{cernrep}
\usepackage{color}
\usepackage{amssymb} 
\usepackage{amsmath}  
\usepackage{slashed}  
\usepackage[normalem]{ulem} 
\usepackage {ulem}
\usepackage{wrapfig}
\begin{document}
\title{Maria Krawczyk :  friend and physicist }
\author{  R.M. Godbole  $^1$ and G. Pancheri $^2 $, \thanks
                 { Also Research Affiliate with MIT CTP, Cambridge, MA, USA}} 
\institute{
$^1$  Center for High Energy Physics, Indian institute of Science, Bangalore, India \\
$^2$ INFN Frascati National Laboratories,  I00044 Frascati, Italy
}
\begin{abstract}
With this brief note, we remember our friend Maria Krawczyk, who passed away this year, on May 24th. We briefly outline some of her physics interests and main accomplishments, and her great human and moral qualities.
\end{abstract}
\maketitle
 \begin{wrapfigure}[35]{r}{0.4\textwidth}
  \begin{center}
  \includegraphics[width=0.4\textwidth]{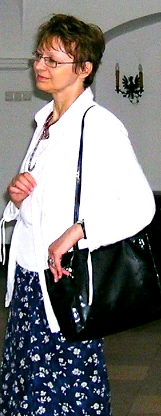}
  \end{center}
\caption{Maria in Kazimierz, 2006.}
\label{fig:maria}
\end{wrapfigure}
On May 25th 2017, the shocking news of Maria Krawczyk's sudden passing reached all the participants  of Photon 2017, the Workshop and Conference on Photons, whose International Advisory Committee  Maria had been a member of  since 2003. Maria had herself organized the 2005 Conference, in Warsaw and Kazimierz, celebrating, at the same time, 100 years since Einstein's paper on the photoelectric effect.

Our friend Maria had passed away the night before,  on May 24th.   All week she had been unwell, but she had thought it was a temporary illness and, until just the  day before,  kept on working with her long time collaborator Ilya Ginzburg. She was also expected, since Monday of that week, to attend the PLANCK 2017 meeting in Warsaw, where she was one of the  Organizers.
 
Maria was a wonderful person to all who knew her and a loving wife, mother and grandmother of four: granddaughter
(15), and 3 grandsons (18, 11 $\&$ 8), to her family.
Her demise is a painful loss for all the people who have known her. First and foremost her family and then a very large number of friends, physics colleagues and collaborators spread all over the world. Her life had not been easy.
As  the communist rule hardened during the 1980's, her husband Tomasz was put in jail, being a strong vocal enemy of the regime. Alone and  fearing  for her husband's future,  she carried  on with her personal and professional life, caring and supporting their  two little boys.  All along,  she was doing  physics, with great passion and a view for unconventional solutions. 

She was deeply interested in a lot of different areas of particle physics and contributed to them very effectively. Her interest in QCD and resummation brought her to Frascati, and to invite  one of us (G.P.) to Warsaw, and Kazimierz Workshops. High energy photons and their hadronic structure \cite{partondis} occupied her for many   years in her working life. She always had a 'nose' for interesting physics and pursued a given area at a time with clear vision of what she wanted to achieve there. She wrote a very useful review on 'Photon Structure Functions' \cite{review}. It was her involvement with the photons that fed her interaction with one  of us (R.G.). Her interest in photons and in the extended Higgs sector, naturally propelled her towards CP studies of the Higgs sector since photon colliders provide one of the most unambiguous probes of the CP property of the Higgs. In fact she was one of the originators of the idea of 'Workshop on CP studies and nonstandard Higgs physics' while she was a scientific associate at CERN.  The idea materialised into a very active and useful workshop \cite{CP}. She also devoted a lot of her mind to the two-Higgs doublet models \cite{doublet}, inert doublet and implications for cosmology thereof \cite{inert}. 

R.G. has known Maria  since 1986, starting from a meeting at the ICHEP conference in Berkeley. At the time Maria was one of the very small number of woman particle theorists, whom one wanted to emulate. She was also the universal mother, caring about everybody with same passion that she cared about physics. She was an incredibly warm  and kind person, as everybody who interacted with her knew.

She was a keen advocate of the future colliders which would be necessary to unravel the physics of the Standard Model and beyond.
She had been involved very deeply in the discussions of the $e^{+}e^{-}$ Linear Collider, first the TESLA being planned in Germany and then the International Linear Collider: ILC. Even closer to her heart was the PLC: Photon Linear Collider in which she got interested during her initial collaboration with Ilya Ginzburg \cite{photoncollider}. 

Her interest  in the subject started in the 90's and led to a long collaboration with Ilya Ginzburg first and then with a number of students as well as some of us. Ilya Ginzburg, was in fact in Warsaw on May 18th, working with her in person until May 20th, and by phone, when she started being unwell.  On May 24th she sent him a  newly corrected version of the paper they are preparing together. So she was immersed in Science till the last moments of her life! Indeed she leaves behind five doctoral students who were working with her. 

With Ilya and Per Osland she had found a corner of parameter space in two-Higgs doublet models 
where a scalar will exist which looks 'like'  the SM Higgs and yet one will not be in the trivial decoupling regime \cite{doubletwithper}. Another long time collaborator of Maria, Per Osland,  always admired her great optimism. One thing Per is keen remembering, and  appreciated very much during the  many years of working with Maria, was her good grasp of experimental facts, and their implications. Due to her interest in the two-Higgs doublet models, Maria was also involved deeply in phenomenology of charged Higgses. Her interest in scalars  led her to co-initiate, with Bohdan Grzadkowski, the very interesting series 'Scalars'  which is held in Warsaw every two years.

Maria was an exceptionally generous soul. Later in her life, after she had become a professor at University of Warsaw, and was recognized as a very good and active particle physicist, after the regime had changed and her husband had been able to return to normal academic life, her mother fell ill and Maria cared for her until the end. In 2006 she organized a {\it Giulia fest} in Kazimierz, for the occasion of the  EURIDICE network Final Meeting, in August 2006. It was the occasion of G.P.'s 65 years of age and the closing meeting of the Research and Training EU Network  of which Maria had been the Polish node scientist-in-charge. Fig.~\ref{fig:maria} shows Maria during  that meeting, in the hall of the Dom Architekta in Kazimierz. The most moving thing in this occasion  is that Maria's mother had just passed away a few days earlier, but Maria courageously kept her promise to be with us and brought the   conference to a very happy and warm ending.

The three of  us, R.G., G.P. and Maria,   shared a special bond, built on  common physics interests, reciprocal respect, and an unspoken  understanding of how difficult it was to charter the path of the career we had chosen. We understood each other, and we will miss her deeply. We are grateful to the friends who helped us in writing this memorial,  in particular to Maria's colleagues and friends Ilya Ginzburg and  Jan Kalinowski,  and to  Maria's  family who shared with us some of their loving memories. 

The short  list of Maria'a papers cited in  this   {\it in memoriam} is certainly incomplete, but we hope it will be sufficient to show her versatility and the depth of physics interests which accompanied her life. 

 \end{document}